\documentclass[runningheads,a4paper]{llncs}

\usepackage{amssymb}
\usepackage{graphicx}
\usepackage{times}
\usepackage{epsfig}
\usepackage{comment}
\usepackage{amsmath}
\usepackage{amssymb}
 \usepackage{booktabs}
 \usepackage{multirow}
\usepackage{pifont}
\usepackage{caption}
\usepackage{subcaption}
\usepackage{xcolor}
\usepackage{float}
\usepackage{url}
\usepackage{soul}
\urldef{\mailsa}\path|{alfred.hofmann, ursula.barth, ingrid.haas, frank.holzwarth,|
\urldef{\mailsb}\path|anna.kramer, leonie.kunz, christine.reiss, nicole.sator,|
\urldef{\mailsc}\path|erika.siebert-cole, peter.strasser, lncs}@springer.com|    
\newcommand{\keywords}[1]{\par\addvspace\baselineskip
\noindent\keywordname\enspace\ignorespaces#1}

\begin{document}

\mainmatter  % start of an individual contribution

% firstathe title is needed
\title{Comparison of automatic prostate
zones segmentation models in MRI images using U-net-like architectures}

%\titlerunning{}

\author{Pablo Cesar Quihui-Rubio\inst{1}, Gilberto Ochoa-Ruiz\inst{1}, Miguel Gonzalez-Mendoza\inst{1},  Gerardo Rodriguez-Hernandez \inst{2} and Christian Mata\inst{3}\inst{,4} }
\authorrunning{Lecture Notes in Computer Science: Authors' Instructions}

\institute{Tecnologico de Monterrey, School of Engineering and Sciences, Mexico. 
\and CIATEQ A.C., Artificial Intelligence Laboratory, Jalisco, 45131, México.
\and Universitat Politècnica de Catalunya, 08019 Barcelona. Catalonia, Spain.
\and Pediatric Computational Imaging Research Group, Hospital Sant Joan de Déu, Esplugues de Llobregat, 08950, Catalonia, Spain \\}


\toctitle{Comparison of automatic prostate
zones segmentation models}
\tocauthor{Authors' Instructions}
\maketitle

%\vspace{2cm}

\begin{abstract}

Prostate cancer is the second-most frequently diagnosed cancer and the sixth leading cause of cancer death in males worldwide. The main problem that specialists face during the diagnosis of prostate cancer is the localization of Regions of Interest (ROI) containing a tumor tissue. Currently, the segmentation of this ROI in most cases is carried out manually by expert doctors, but the procedure is plagued with low detection rates (of about 27-44\%) or over-diagnosis in some patients. Therefore, several research works have tackled the challenge of  automatically segmenting and extracting features of the ROI from magnetic resonance images, as this process can greatly facilitate many diagnostic and therapeutic applications. However, the lack of clear prostate boundaries, the heterogeneity inherent to the prostate tissue, and the variety of prostate shapes makes this process very difficult to automate.In this work, six deep learning models were trained and analyzed with a dataset of MRI images obtained from the Centre Hospitalaire de Dijon and Universitat Politecnica de Catalunya. We carried out a comparison of multiple deep learning models (i.e. U-Net, Attention U-Net, Dense-UNet, Attention Dense-UNet, R2U-Net, and Attention R2U-Net) using categorical cross-entropy loss function. The analysis was performed using three metrics commonly used for image segmentation: Dice score, Jaccard index, and mean squared error. The model that give us the best result segmenting all the zones was R2U-Net, which achieved 0.869, 0.782, and 0.00013 for Dice, Jaccard and mean squared error, respectively.

\end{abstract}

\section{Introduction}
\label{introduction}

%Relevance of pca and general idea of biomedical imaging
Prostate cancer (PCa) is the second leading cause of cancer deaths in the world and nowadays one of eight men are diagnosed with this disease in their lifetime \cite{ACS}. There are some risk factors, such as the age above 50 years, family history, obesity, ethnicity that must be considered during the diagnosis process, and it is noteworthy that the survival rate for regional PCa is almost 100\% when detected in early stages. In stark contrast, the survival rate when the cancer is spread to other parts of the body is of only 30\% \cite{astrazeneca_2020}.

\begin{comment}

Current methods for early detection of  PCa are through a transrectal ultrasound (TRUS) guided biopsy, which consists in taking a sample of tissue from the prostate using a needle with an ultrasound scanner to examine in the laboratory. However, medical imaging (magnetic resonance imaging (MRI), ultrasound (US), and computed tomography (CT)) represent important non-invasive tools for aiding in the prevention, detection, and monitoring of different types of cancer, including PCa \cite{aldoj2020}.
\end{comment}

Magnetic Resonance Imaging (MRI) has been established as the best medical image tool for the detection, localization and staging of PCa, due to their high resolution, excellent spontaneous contrast of soft tissues, and the possibility of multi-planar and multi-parametric scanning \cite{Chen_2008}. Although MRI has been traditionally used for staging PCa, we will focus on PCa detection trough ROI segmented from MR images. 

%Prostate segmentation relevance
The use of image segmentation of MR images for PCa detection and characterisation can in fact help in determining the tissue volume, aiding as well in the localization the cancerous tissue in the ROI \cite{Haralick_1985}. Thus, an accurate and consistent segmentation is crucial in PCa. Although prostate segmentation is a relatively old problem and some methods have been proposed in the past using conventional image processing pipelines, nowadays, the most common and traditional method to identify and delimit prostate gland and prostate regions of interest (central zone, peripheral zone, transition zone) is performed manually by radiologists \cite{aldoj2020}. 

This non-automated process has been proven to be time-consuming and, due to the subjectivity of the task and different interpretations from multiple specialists, it is highly operator dependant and difficult to reproduce \cite{Rasch_2011}. Therefore, automating this process for the segmentation of prostate gland and regions of interest, in addition to saving time for radiologists, can be used as a learning tool for others and have consistency in contouring \cite{Mahapatra_2014}.

\begin{comment}
\par Nevertheless, segmentation in MR images is a complicated task for many reasons. According to Aldoj et \textit{al.} \cite{aldoj2020}, there are three main reasons why the segmentation of central zone (CZ), peripheral zone (PZ) and transition zone (TZ) are difficult using MRI. First, the boundaries of the regions are very diffused, thus, is difficult to identify the gland from other no relevant tissues, and it is common to get an over-segmentation. The second reason is the variability of MRI techniques and machines, getting different signal intensity on the images. Third, there are multiple shapes, sizes and tissue types between patients or the appearance of a pathology.
\end{comment}

%Relevance of deep learning
\par In recent years, the automatic segmentation of the prostate has been promoted with the use of deep learning techniques. These methods, called convolutional neural networks (CNNs) have generated results that outperform traditional methods due to their ability to learn complex features and perform an accurate classification of pixels, resulting in segmentation \cite{ELGUINDI_2019}. Several works have addressed the problem of prostate segmentation using deep learning, such is the case of the popular U-Net \cite{Unet_ronneberger_2015} model, which is the base of many recent works in literature: \textit{MultiResU-Net} \cite{multires2020}, \textit{Dense-UNet} \cite{dense_unet_wu_2021}, \textit{Attention U-Net} \cite{Att-unet-2018}, among others. Although good results have been obtained by the authors of these models, there is a lack of datasets with enough information to segment the prostate and all of its ROI correctly.

\par In this paper, we explore six recent deep learning methods (i.e. U-Net, Attention U-Net, Dense-UNet, Attention Dense-UNet, Recurrent Residual Convolutional Neural Network based on U-Net (R2U-Net), and Attention R2U-Net architectures.) to segment the prostate and evaluate their performance in this task. Even though a comparison of metrics to evaluate the performance between these models have been performed in other works, most of them only evaluate between the whole prostate gland or two principal zones: CZ, and PZ.  

\begin{figure}
    \centering
    \includegraphics[width = 1\textwidth
    ]{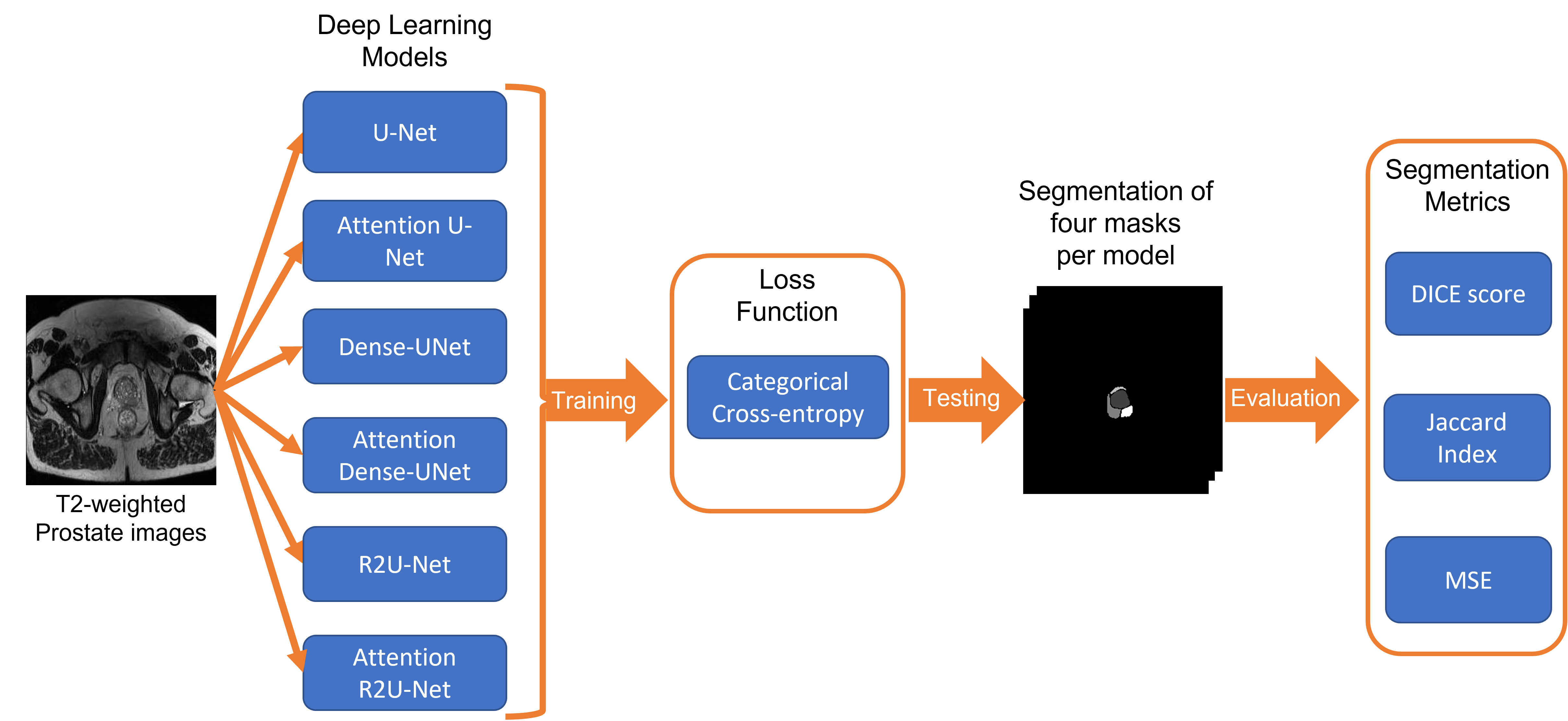}
    
    \caption{Pipeline of experiments. Six models were trained with 16 patients (205 images) using four loss functions. As a result, an image with four segmented zones was obtained for each test image on all models. The resulting images were evaluated using three segmentation metrics.}
    \label{fig: pipeline}
\end{figure}

The dataset we used to perform our experiments the Centre Hospitalaire de Dijon and consists of 16 patients, with a total of 205 images with their corresponding annotations that were validated by a collaboration of experts using a dedicated software tool \cite{Mata_2020}. The deep learning models studied and compared were trained in the same conditions, using categorical cross-entropy loss function and three metrics to measure their performance. In this work, we make a thorough comparison between six models, including five U-Net based models and original U-Net, using multiple loss functions and the same metrics for evaluation. The pipeline of our experiments is shown in Figure \ref{fig: pipeline}.

%Pipeline of the experiments

%Content pf the paper

This paper has five sections including this introduction. Section \ref{State_of_the_art} is divided in two subsections where we mentioned the motivation of doing this study, and also, discuss previous works related to prostate segmentation, focusing on deep learning methods. Section \ref{data_methods} is divided in four subsections, where we described the dataset, deep learning architectures, metrics, loss functions, and details for training and testing. In section \ref{results} the results of the experiments are discussed in detail. The conclusion and future work are in Section \ref{conclusions_future_work}.

\section{Motivation and State of the art}
\label{State_of_the_art}

\subsection{Motivation for segmenting prostate zones}
\par In accordance with Sun et \textit{al}. \cite{sun_2019}, analyzing multiparametric MRI (mpMRI) images is a technique used in patients with possible PCa, which can be performed before a transrectal ultrasound (TRUS) or after a negative TRUS. In the case of a potential cancerous tissue, it can also be analyzed by MRI guided biopsy or MRI-US guided biopsy, which both of them have shown higher accuracy than only using TRUS \cite{sun_2019}. For patients who have been correctly diagnosed with PCa, morphology and localization are features that can be extracted from mpMRI images. An accurate localization of tumor can be carried out by mpMRI, chiefly those in anterior zone that TRUS may miss \cite{Gupta_2013}. Also, mpMRI has shown a high accuracy defining tumor volume, which is a risk factor \cite{sun_2019}. 

\par Therefore, segmentation of MRI images is crucial to define the prostate boundaries (including zones) and exclude other nearby organs that are not of interest at the moment \cite{sun_2019}. As commented before, there are manual and automated methods to do it and the most common used is manual. 

This manual method has several limitations that can affect in the analysis of PCa detection, such as time consumption, subjective results, variability, etc., and, for that reason, automated techniques are currently the main discussion in research \cite{aldoj2020}.The difficulty in the segmentation process is due to the complex nature of medical images and the presence of non-linear features in most of them \cite{sharma2010}. 

More precisely, the segmentation output can be affected by: intensity inhomogeneity, closeness in gray level of different soft tissue, partial volume effect, or presence of artifacts \cite{sharma2010}, which are not easy to fix or homogenize between acquisition or among patients and are highly operator dependant as well.

Another issue in image segmentation pertains validation fot the results by several specialists  following a reproducible method. New automated methods have tried to overcome these limitations, and there are some examples such as U-net \cite{Unet_ronneberger_2015} network architecture which is dedicated to segment biomedical images, but it has some drawbacks \cite{aldoj2020}. 

\subsection{Related Work}

\subsubsection{Machine Learning methods}

%atlas-based models, deformable models, feature-based models

Several traditional methods in the literature for prostate zones segmentation have been proposed. They can be classified as atlas-based models, deformable models, and feature-based ML methods \cite{Li_2022}. Atlas-based models consist in a collection of multiple images segmented manually by experts, which are used as a reference for new segmentation in images of other patients \cite{Li_2022}. A study of Klein et \textit{al.} \cite{Klein_2007} in 2007, proposed a model based in MR atlas images, they registered the target image in a non-rigidly way using a similarity measure, to use the same measure to select the best ones, and obtain the segmentation by averaging the selected deformed segmentation. They obtained a median Dice similarity coefficient (DSC) of 0.82 evaluating their model in 22 images \cite{Klein_2007}. 

Deformable models is another technique that has been used in the literature to get an accurate prostate segmentation, this models are based in mathematical, geometrical, and physical theories to constrain and guide a curve to delineate an object's border \cite{reda_2016}. Liu et \textit{al.} \cite{Liu_2011} in 200 proposed a deformable model using level set in MRI data, the model was tested in images from 10 patients and they obtained a DSC score of 0.81. 

Another methods were introduced in order to get a more precise results of segmentation in medical devices such as: k-means clustering, thresholding, active contour methods, among others. Those techniques have shown good results in image segmentation. However, in the last years, the field of Deep Learning (DL) has growth exponentially in medical imaging, particularly in the segmentation process \cite{comelli_2021}.

\subsubsection{Deep Learning based methods}

\par In medical imaging, deep-learning techniques have been improving the analysis, such as image segmentation, image registration, image fusion, image annotation, CADx, lesion detection, and microscopic image analysis \cite{Shen_2017}. In recent years, new DL models for image segmentation have been developed, specially focused on the biomedical area. 

One of the best known models in the literature is U-Net, which is a CNN composed of a series of four convolution and max-pooling operations which reduce the dimension of the input image, followed by four convolution and up-sampling operations \cite{Unet_ronneberger_2015}.
\par There are some works that use U-Net to get an automatic segmentation of the prostate. Zhu et \textit{al.} \cite{Zhu_deeplycnn_2017} proposed a deeply-supervised CNN to segment the whole prostate gland, based on the structure of U-Net getting a mean DSC of 0.885. Also, Zabihollahy et \textit{al.} \cite{zabihollahy2019} designed a model composed by two U-Net architectures to segment the prostate gland, as well as, its CZ and PZ in MRI T2-weighted images and ADC maps, obtaining a mean DSC score of 0.92, 0.91, 0.86, respectively. Also, Clark et \textit{al.} \cite{clark_2017} presented a new architecture based on U-Net and inception model to segment the prostate and transition zones using diffusion-weighted MR images, they obtained a mean DSC of 0.93 and 0.88, respectively. Rundo et \textit{al.} \cite{Rundo_usenet_2019} proposed a novel architecture called USE-Net, which incorporates Squeeze-and-Excitation into U-Net; they achieved a segmentation of the prostate zones outperforming most of the state-of-the-art results in peripheral zone segmentation. In another work, Runo et \textit{al.} \cite{Rundo_datasets_2019} analyzed some CNN models with datasets from different institutions using only T2-weighted MR images and concluded that U-Net outperforms other methods in the state of the art. More recently, Aldoj et \textit{al.} \cite{aldoj2020} proposed a novel model based on U-Net and DenseNet and did a segmentation of prostate zones in 3D MR images with three variations of their architecture, they obtained a mean DSC for the whole gland, CZ, and PZ of 0.92, 0.89, and 0.78, respectively.  

%Attention gates
\par In 2018, Oktay et \textit{al.} \cite{Att-unet-2018} proposed a novel attention gates to incorporate it into the existing U-Net model. The intention of using attention blocks is that, in an automatic way, the model learns to focus on the specific target structures, and ignore the rest of them on the image. In Attention U-Net model, the attention gates highlight the salient features from the skip connections between the encoder and decoder \cite{Att-unet-2018}. These attention gates modules have been implemented in other architectures such as Attention Dense-UNet \cite{Att-denseunet-2019}, Spatial Attention U-Net \cite{SAUNetSA}, Attention R2U-Net, among others.

\par Although automatic prostate segmentation has improved during the last years, there is still work to do in segmentation of specific prostate zones such as CZ, PZ, and TZ. An accurate segmentation of these zones, as well as, of tumor (TUM) of different sizes and shapes, could lead in early detection of prostate cancer. Therefore in this work, we compared some models from the literature using categorical cross-entropy as loss function with a dataset of only T2-weighted images. We analyzed the segmentation of the prostate zones using different metrics to choose the best DL architecture.

\section{Data and methods}
\label{data_methods}

The technical contribution of this work is the evaluation of the impact of four loss functions with three metrics on the prostate segmentation using U-Net, Attention U-Net, Recurrent Residual Convolutional Neural Network based on U-Net (R2U-Net), Attention R2U-Net, , Dense-UNet, and Attention Dense-UNet architectures. A total of 6 segmentation processes with a maximum of four zones per image are compared in this work. A visual summary of the experiments carried out is shown in Figure \ref{fig: pipeline}. In this section the data and methodology followed is described and explained.

\subsection{Dataset}
\label{Dataset}

The dataset of images used in this work were provided by PhD. Christian Mata from UPC in Barcelona in collaboration with DIJON hospital in France. The examinations used in our study contained three-dimensional T2-weighted fast spin-echo (TR/TE/ETL: 3600 ms/143 ms/109, slice thickness: $1.25$ mm) images acquired with sub-millimetric pixel resolution in an oblique axial plane..The institutional committee on human research approved the study, with a waiver for the requirement for written consent, because MRI was included in the workup procedure for all patients referred for brachytherapy or radiotherapy.

In addition to the images, a manual segmentation of each with four regions of interest (CZ, PZ, TZ, and TUMOR) was provided and this process was cautiously validated by multiple professional radiologists and experts using a dedicated software tool \cite{Mata_2020}. The format of the T2-weighted scans from the dataset are DICOM, with their correspondent annotations in format CSV of the prostate zones.

% \hl{The data used in this work is conformed of MR prostate images obtained by \hl{Hospital Dijon} in collaboration with PhD. Christian Mata from Universitat Politecnica de Catalunya.} The images are of type T2-weighted captured with a MRI machine of \hl{3} Teslas and were taken from \hl{19} patients with a total of \hl{X} images. The format of the T2-weighted scans from the dataset are DICOM, with their correspondent annotations in format CSV of four prostate zones: CZ, PZ, TZ and Tumor. 

\par
\begin{figure}[b]
    \centering % <-- added
    \begin{subfigure}{0.5\textwidth}
      \includegraphics[width=0.6\textwidth]{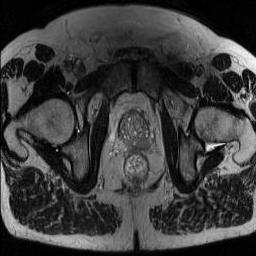}
      \centering
      \caption{Original T2-MRI prostate image}
      \label{fig: prostate_Img}
    \end{subfigure}\hfil % <-- added
    \begin{subfigure}{0.5\textwidth}
      \includegraphics[width=.6\textwidth]{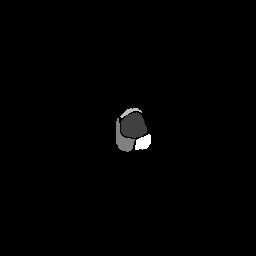}
      \centering
      \caption{Masks of prostate zones}
      \label{fig:prostate_label}
    \end{subfigure}\hfil % <-- added

\caption{Sample image and mask of the dataset.}
\label{fig:dataset_ex}
\end{figure}

 The ground truth masks were created from the CSV data files in a resolution of 256x256 pixels. Also, DICOM files were transformed to images of the same resolution after a data augmentation process was carried out. As is common with medical images, due to the difficulty of obtaining good images to work with them, a biomedical image augmentation algorithm \cite{augmentor} was performed. This process help us to increase the number of training images, and consisted of 16 geometrical transformations such as: rotations, zoom, translations, among others variations. At the end, the total number of images with their corresponding masks were 3485. An example from the dataset is shown in Figure \ref{fig:dataset_ex}; in the image there is an original T2-MRI image of the prostate and the mask of the zones generated from the data file provided.

\subsection{Deep Learning Architectures}
\label{tested models}
As mentioned in Section \ref{State_of_the_art}, there are several deep learning architectures used for image segmentation in the literature. In this work, we focused on five models based in the U-Net architecture originally proposed by Ronnerberget et \textit{al.} \cite{Unet_ronneberger_2015}, and this one was also considered for comparison. 

\par One of the architectures used in this work is Attention U-Net by Oktay et \textit{al.} \cite{Att-unet-2018}, which incorporates Attention Gates (AGs) into the standard U-Net architecture to highlight salient features that are passed through the skip connections. This is performed right before the concatenation operation to merge only relevant activations \cite{Att-unet-2018}. AGs progressively suppress feature responses in irrelevant background regions without the requirement to crop a regions of interest between networks \cite{Att-unet-2018}.
\par Also, Dense-UNet proposed by Wu et \textit{al.} \cite{dense_unet_wu_2021} was selected for the comparison in this work, which consists of a network that combines the U-Net architecture with dense concatenation to reduce resolution loss. The network consists of a right side of the architecture with a dense downsampling path, and the left side with a dense upsampling path, also, it incorporates some skip connection channels to connect the paths.
\par Other model tested was R2U-Net presented by Zahangir et \textit{al.} \cite{r2unet} consists of a recurrent residual convolutional neural network which has been demonstrated that outperforms classical U-Net due to the benefit of feature accumulation inside the model in training and testing processes, among other novel features proposed.
\par Attention Dense-Unet proposed by Li et \textit{al.} \cite{Att-denseunet-2019}, is an integration of Attention modules and the model Dense-UNet, and has been demonstrated in the literature to outperforms Dense-UNet, thus we decided to include it in the comparison. Finally, we did an integration of Attention U-Net \cite{AttUnet} and R2U-Net \cite{r2unet} architectures called Attention R2U-Net to get a combination of the benefits of both models and compare the performance of the segmentation tasks.

\subsection{Segmentation metrics}
\label{Segmenatation metrics}
There are several metrics used for image segmentation in the literature \cite{taha2015} and the selection of metrics for evaluation depends on the data and segmentation task. Therefore, the metrics we have selected to be used on this work aiming to get a robust comparison between the segmentation architectures and loss functions are Dice Similarity Coefficient (DSC), Jaccard Index (JAC), and Mean Square Error (MSE). 
\par As  are trying to segment multiple prostate zones, it is necessary to use the appropriate metrics, as mentioned by Taha et \textit{al.} \cite{taha2015}, the use of Jaccard index and DSC is appropriate in this case. The Jaccard Index, also known as Intersection Over Union (IOU), is a segmentation metric based on overlap and it is defined as the intersection between ground-truths and predictions divided by their union (see Equation \ref{eqn: JAC}). DSC is the most used metric based for calculating the overlap between the ground-truth and predicted images divided by the common pixels between them, and it can be defined as shown in equation \ref{eqn: DICE}. The last metric is Mean Square Error (MSE), which averages the difference between the ground-truths and predicted segmentation.

\begin{equation}
\text{JAC} = \frac{|\text{Prediction}\: \cap\: \text{Ground Truth}|}{|\text{Prediction} \cup \text{Ground Truth}|}
\label{eqn: JAC}
\end{equation}

\begin{equation}
\text{DSC} = \frac{2|\text{Prediction}\: \cap\: \text{Ground Truth}|}{|\text{Prediction}| + |\text{Ground Truth}|}
\label{eqn: DICE}
\end{equation}

\subsection{Loss functions}
\label{loss_functions}

The choice of a loss function is extremely important for any deep learning architecture, due to the fact that it guides the learning process of the algorithm. That is to say, it makes the algorithm to be more accurate, faster and reproducible during the training process. Also, a correct selection of loss function can reduce or mitigate the problem of overfitting in the model. Herein, we evaluated the models with Categorical Cross-Entropy (CCE), which is a common loss function used in multi-class segmentation, and it is designed to quantify the difference between two probability distributions \cite{Yeung_2021}.

\subsection{Training}

All the models in this work were trained using equal parameters and settings; the five tested models were implemented using Keras. The U-Net model was implemented using an adapted code for multiclass segmentation provided by Sha, Y. \cite{keras-unet-collection}. The implementation of Dense-UNet made by Wu et \textit{al.} \cite{dense_unet_wu_2021} was used in this work. For the Attention U-Net, R2U-Net, and Attention R2U-Net, the codes implementations were taken from a Github repository \cite{pytorch_implementation} and transformed to be used with Keras. Finally, Attention Dense-UNet architecture was designed combining Dense-UNet and Attention U-Net implementation.
\par The training process for each model was performed using 90\% of images, 100 epochs, a batch size of 6, and a learning rate of 0.0001. All the training was done using a NVIDIA DGX workstation, using two V100 GPUs for each model. At the end, the trained weights were saved for future testing and prediction.

\section{Results}
\label{results}

As mentioned above, previous works in the literature that have shown great promise in prostate segmentation. However, most of them have focused solely on specific zones. Segmenting the gland or only CZ and PZ, is a more common task due to its boundaries are more delimited than zones such as TZ and Tumor, where the size and shape presents more variability between patients and images. Therefore, comparing multiple deep learning architectures dedicated for medical image segmentation with a dataset with images and masks of all the zones of the prostate represents an interesting contribution. 

In what follows, we will perform a comparison of the obtained results both quantitatively (through the selected metrics) and qualitatively (by comparing various sample images and their corresponding GT) to assess the performance of studied models.

\subsection{Quantitative Results}

In Table \ref{tbl: Metrics results} shows a summary of the five models performance measured under the metrics mentioned before. A  ($\uparrow$)  or lower ($\downarrow$) symbol indicates whether this metrics has to be maximized or minimized for obtaining a better model, whereas the values correspond to the mean metric between the prostate zones and test images. The bold values represents the model that achieved the best metric score within all of them.

\renewcommand{\arraystretch}{1}
\begin{table}[t]
\setlength{\tabcolsep}{18pt}
\caption{Results using multiple deep learning models and metrics (average) for Categorical Cross-Entropy Loss.}
\label{tbl: Metrics results}
\centering
\begin{tabular}{lcll}
\hline
\multirow{2}{*}{\textbf{Model}}&\multicolumn{3}{c}{Metrics}\\\cline{2-4}
&\textbf{DSC $\uparrow$}&\textbf{JAC $\uparrow$}&\textbf{MSE $\downarrow$}\\
\hline
% \textbf{Model} & \multicolumn{1}{c}{\textbf{DSC $\uparrow$}} & \multicolumn{1}{c}{\textbf{JAC $\uparrow$}} & \multicolumn{1}{c}{\textbf{MSE $\downarrow$}} \\ \hline
U-Net & 0.731 & 0.635 & 0.0021 \\ %\cline{2-4} 
Attention U-Net & 0.839 & 0.741 & 0.0016 \\ %\cline{2-4} 
Dense-UNet & 0.830 & 0.725 & 0.0018 \\% \cline{2-4} 
Attention Dense-UNet & 0.844 & 0.747 & 0.0016 \\ %\cline{2-4} 
R2U-Net & \textbf{0.869} & \textbf{0.782} & \textbf{0.0013} \\ %\cline{2-4} 
Attention R2U-Net & 0.864 & 0.775 & 0.0014 \\ \hline
\end{tabular}

\end{table}

% Analysis of results by models
Although all the models analyzed in this work are U-Net based, with our dataset this base architecture had the worst results in metrics performance. After incorporating attention modules to U-Net, the performance increased  in 0.108, 0.106, and 0.05 for the DSC, JAC, and MSE metrics, respectively. On the other hand, Dense-UNet model achieved lower metric values before integrating  attention gates. After integrating attention, the performance obtained by Attention Dense-UNet was the best. 

Analyzing all the results in Table \ref{tbl: Metrics results}, it is can be inferred that R2U-Net is the model with the highest performance in Dice and Jaccard metrics (although a lowest value for MSE was obtained). The gain in performance between R2U-Net and the base U-Net model is around 13.8\%, 14.7\%, and 0.0008 for DSC, JAC, and MSE, respectively. However, the architecture of R2U-Net with Attention modules obtained almost the same values in the test set, with a difference of only 0.004, 0.007 and 0.0001, respectively. 

\begin{figure}[t]
    \centering
      \includegraphics[width=\linewidth]{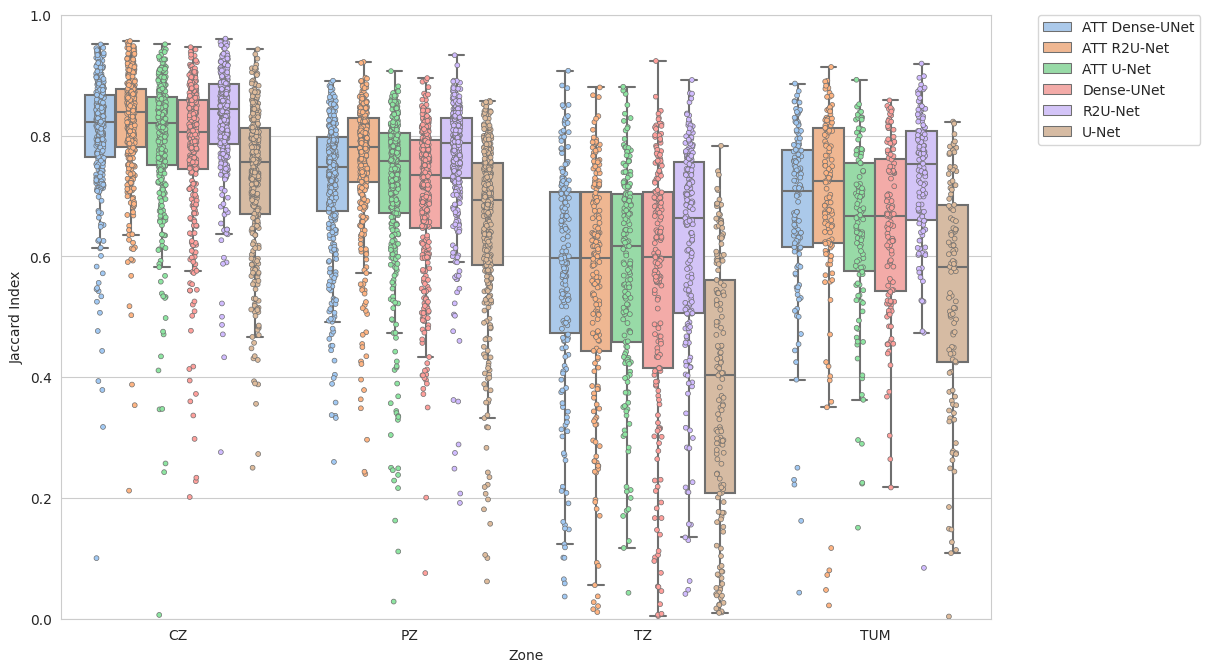}
    \caption{Comparison of Jaccard Index per each class between Deep Learning Architectures. Each predicted segmentation is represented by a colored dot to visualize the variation in the results of the models.}
    \label{fig: iou_classes}
\end{figure}

However, the two last models require more computational resources (and inference time). The difference in performance can be explained by the use of attention blocks, which requires extra parameters, possibly leading to overfitting. Nevertheless, due to the little discrepancy between the metrics in these two architectures, it could be difficult to notice the gap in a visual comparison with images. For comparing the segmentation performance of the models for each zone (CZ, PZ, TZ, and TUMOR), we selected the Jaccard Index metric ($\uparrow$) for evaluation. A summary of the results is shown in a box-plot (see Figure \ref{fig: iou_classes}), were each color represents a different model, and each point is the result of the Jaccard performance in a tested image.

 %comparison by zones

As is shown in Figure \ref{fig: iou_classes}, 
the performance of the models varies according to the prostate zone segmented. The zone with the best Jaccard index, aside from the background, was the CZ, where all of the models performed similarly, with low variance between them. However, the best model was R2U-Net with a JAC value of 0.84, and the worst performance in this zone was U-Net base model.

\par For the segmentation of the peripheral zone, all the models behaved similarly, but they exhibited lower values for the metrics of interest in average. Although the other 2 zones were segmented with high values of JAC, it can be seen on Figure \ref{fig: iou_classes} that the variance in this zone is higher than in CZ, yielding much lower values in some models such as U-Net and Dense-UNet. In this zone, the models with lower variance and higher mean score are R2U-Net and Attention R2U-Net with mean JAC index of 0.763 and 0.751, respectively.

\par The Transition zone (TZ) was the one with worst segmentation performance for all the models, with high variances and an average JAC index below 0.57, achieved by the best architecture, R2U-Net. The variance and poor performance in this region can be explained by overall lack of ground truth masks of this zone in the training data, as well for the relatively small size of the zone compared to others. This zone is difficult to segment even for radiologists, so it is commonly taken as part of CZ or PZ. The above-mentioned problems could be solved by including more patients, with better delimited TZ masks to our dataset. Finally, the tumor was the last region analyzed, with less dispersion of points for all the models, but there was still variance for a few sample images. As in the previous zones, the best models form the tumor class were R2U-Net and Attention R2U-Net, with an average JAC index of 0.74 and 0.71, respectively. U-Net and Dense-UNet were the worst models; however, the implementation of attention modules improved the performance of those models by 11.6\% and 39.3\%, respectively. 

\subsection{Qualitative Results}

\begin{figure}[t]
    \centering
    \includegraphics[width = 0.8\textwidth]{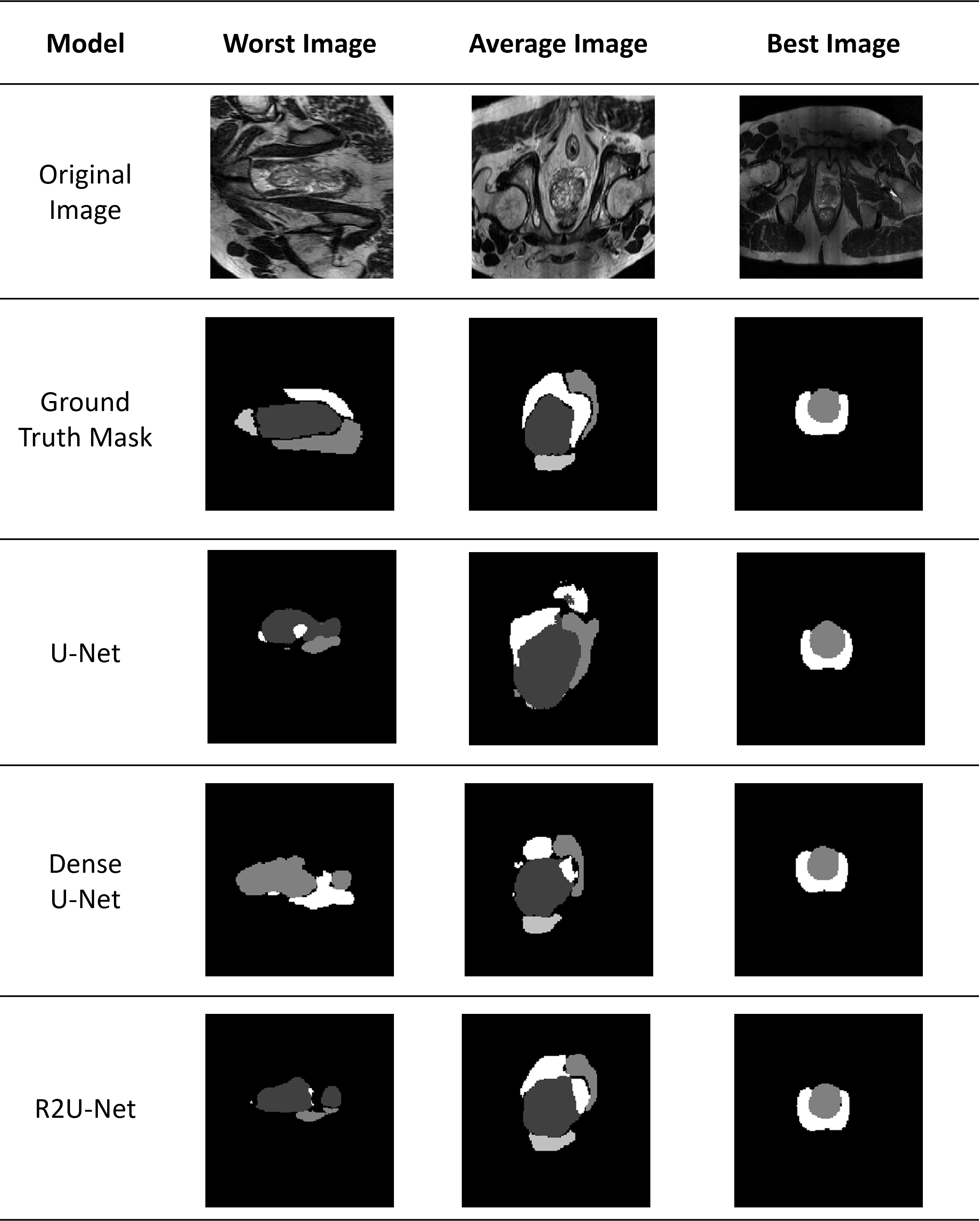}
     \caption{Comparison of the worst, average and best images (obtained from R2U-Net) predicted by the worst, average, and best architectures analyzed in this work.}
    \label{fig: pred}
\end{figure}

Figure \ref{fig: pred} presents a qualitative comparison for the tested models against the ground truth. We have selected three examples for assessing which models perform the best in different scenarios. The images were selected  based on the JAC values obtained by R2U-Net (our best performing model) with the idea of showing the mos representative examples of the worst case image (first column), the image with the average JAC value in the dataset (center column) and the best segmented image in the entire test set (third column). The first row shows the input image, while the second depicts the GT mask, and each subsequent row displays the generated segmentation for each input image

As expected, the R2U-Net produces very good results for all the zones in a great share of the examples, but it struggles with the tumor zone and the transition zone, as the quantitative results suggested. Surprisingly,  the U-Net model, which was the worst one, had acceptable results in images with only two or three zones to be segmented, as well as, Dense U-Net model.

\section{Conclusion and future work}
\label{conclusions_future_work}

After the analysis between all the models using the same conditions, the best architecture to segment prostate and its zones was R2U-Net. The segmentation of all the zones in any T2-weighted MR image is not easy, and in many cases there is not possible to delimit the boundaries of transition zone, even for experts. 

This could be one of the reasons this zone is so difficult to segment for all the tested models. Even though all the models are U-Net based, there were differences in performance between them during segmentation tasks. The incorporation of attention gates in the U-Net and Dense-UNet architectures yielded better average metrics values, but  this was not the case for Attention R2U-Net architecture. This could be due to the huge increase in the number of parameters in the A-R2U-Net model, which led to overfitting sooner that other models. Nevertheless, the difference between Attention R2U-Net and only R2U-Net metrics in average is not significant, but due to the resources and time consumption, we decided that the best model for training is R2U-Net.

It is well-known that the segmentation of prostate zones is a difficult task, even for radiologists, but it is possible to achieve good results using an automatic model with a verified dataset and the correct selection and implementation of a deep learning architecture for training. In this study, we concluded that the best option is R2U-Net, however there is still work to do in some zones such as TZ and Tumor, where it is desired to get higher dice score and Jaccard index. In order to achieve that, as a future work, we should train models with a larger dataset, as well as trying different loss functions focused on imbalanced datasets, which can also reduce the variance in our results. 

\vspace{-0.3cm}
\section*{Acknowledgments}
\vspace{-0.1cm}
The authors wish to thank the AI Hub and the CIIOT at ITESM for their support for carrying the experiments reported in this paper in their NVIDIA's DGX computer.

\bibliographystyle{unsrt}
\bibliography{References}

\begin{thebibliography}{10}

\bibitem{ACS}
American~Cancer Society.
\newblock Key statistics for prostate cancer: Prostate cancer facts.
\newblock
  \url{https://www.cancer.org/cancer/prostate-cancer/about/key-statistics.html}.
\newblock Accessed October 17, 2021.

\bibitem{astrazeneca_2020}
AstraZeneca.
\newblock A personalized approach in prostate cancer.
\newblock
  \url{https://www.astrazeneca.com/our-therapy-areas/oncology/prostate-cancer.html},
  May 2020.
\newblock Accessed October 17, 2021.

\bibitem{Chen_2008}
M.~Chen, H-D. Dang, J-Y. Wang, C.~Zhou, S-Y. Li, W-C. Wang, W-F. Zhao, Z-H.
  Yang, C-Y. Zhong, and G-Z. Li.
\newblock Prostate cancer detection: Comparison of t2-weighted imaging,
  diffusion-weighted imaging, proton magnetic resonance spectroscopic imaging,
  and the three techniques combined.
\newblock {\em Acta Radiologica}, 49(5):602--610, 2008.

\bibitem{Haralick_1985}
R.~Haralick and L.~Shapiro.
\newblock Image segmentation techniques.
\newblock {\em Computer Vision, Graphics, and Image Processing},
  29(1):100--132, 1985.

\bibitem{aldoj2020}
N.~Aldoj, F.~Biavati, F.~Michallek, S.~Stober, and M.~Dewey.
\newblock Automatic prostate and prostate zones segmentation of magnetic
  resonance images using densenet-like u-net.
\newblock {\em Scientific Reports}, 10, 08 2020.

\bibitem{Rasch_2011}
{Coen R.N.} Rasch, {Joop C.} Duppen, {Roel J.} Steenbakkers, Daniel Baseman,
  {Tony Y.} Eng, {Clifton D.} Fuller, {Anna M.} Harris, {William E.} Jones,
  Ying Li, Elizabeth Maani, {Dominic D.} Nguyen, {Gregory P.} Swanson, Celine
  Bicquart, Patrick Gagnon, John Holland, Tasha McDonald, {Charles R.} Thomas,
  Samuel Wang, Martin Fuss, {Hadley J.} Sharp, Michelle Ludwig, {David I.}
  Rosenthal, {Aidnag Z.} Diaz, {Carlo G.N.} Demandante, and Ronald Shapiro.
\newblock Human-computer interaction in radiotherapy target volume delineation:
  A prospective, multi-institutional comparison of user input devices.
\newblock {\em Journal of Digital Imaging}, 24(5):794--803, October 2011.

\bibitem{Mahapatra_2014}
D.~Mahapatra and J-M. Buhmann.
\newblock Prostate mri segmentation using learned semantic knowledge and graph
  cuts.
\newblock {\em IEEE Transactions on Biomedical Engineering}, 61(3):756--764,
  2014.

\bibitem{ELGUINDI_2019}
Sharif Elguindi, Michael~J. Zelefsky, Jue Jiang, Harini Veeraraghavan,
  Joseph~O. Deasy, Margie~A. Hunt, and Neelam Tyagi.
\newblock Deep learning-based auto-segmentation of targets and organs-at-risk
  for magnetic resonance imaging only planning of prostate radiotherapy.
\newblock {\em Physics and Imaging in Radiation Oncology}, 12:80--86, 2019.

\bibitem{Unet_ronneberger_2015}
Olaf Ronneberger, Philipp Fischer, and Thomas Brox.
\newblock U-net: Convolutional networks for biomedical image segmentation.
\newblock In Nassir Navab, Joachim Hornegger, William~M. Wells, and
  Alejandro~F. Frangi, editors, {\em Medical Image Computing and
  Computer-Assisted Intervention -- MICCAI 2015}, pages 234--241, Cham, 2015.
  Springer International Publishing.

\bibitem{multires2020}
Nabil Ibtehaz and M.~Sohel Rahman.
\newblock Multiresunet : Rethinking the u-net architecture for multimodal
  biomedical image segmentation.
\newblock {\em Neural Networks}, 121:74--87, 2020.

\bibitem{dense_unet_wu_2021}
Yufeng Wu, Jiachen Wu, Shangzhong Jin, Liangcai Cao, and Guofan Jin.
\newblock Dense-u-net: Dense encoder–decoder network for holographic imaging
  of 3d particle fields.
\newblock {\em Optics Communications}, 493:126970, 2021.

\bibitem{Att-unet-2018}
Ozan Oktay, Jo~Schlemper, Loic~Le Folgoc, Matthew Lee, Mattias Heinrich,
  Kazunari Misawa, Kensaku Mori, Steven McDonagh, Nils~Y Hammerla, Bernhard
  Kainz, Ben Glocker, and Daniel Rueckert.
\newblock Attention u-net: Learning where to look for the pancreas, 2018.

\bibitem{Mata_2020}
J.~Rodríguez, G.~Ochoa-Ruiz, and C.~Mata.
\newblock A prostate mri segmentation tool based on active contour models using
  a gradient vector flow.
\newblock {\em Applied Sciences}, 10(18), 2020.

\bibitem{sun_2019}
Y.~Sun, H-M. Reynolds, B.~Parameswaran, D.~Wraith, M-E. Finnegan, S.~Williams,
  and A.~Haworth.
\newblock Multiparametric mri and radiomics in prostate cancer: a review.
\newblock {\em Australasian physical \& engineering sciences in medicine},
  42(1):3--25, 2019.

\bibitem{Gupta_2013}
R.~Gupta, C.~Kauffman, T.~Polascik, S.~Taneja, and A.~Rosenkrantz.
\newblock The state of prostate mri in 2013.
\newblock {\em Oncology (Williston Park)}, 27(4):262--70, 2013.

\bibitem{sharma2010}
Neeraj Sharma and Lalit~M Aggarwal.
\newblock Automated medical image segmentation techniques.
\newblock {\em Journal of medical physics/Association of Medical Physicists of
  India}, 35(1):3, 2010.

\bibitem{Li_2022}
Huanye Li, Chau~Hung Lee, David Chia, Zhiping Lin, Weimin Huang, and Cher~Heng
  Tan.
\newblock Machine learning in prostate mri for prostate cancer: Current status
  and future opportunities.
\newblock {\em Diagnostics}, 12(2), 2022.

\bibitem{Klein_2007}
S.~Klein, U.A. van~der Heide, B.W. Raaymakers, A.N.T.J. Kotte, M.~Staring, and
  J.P.W. Pluim.
\newblock Segmentation of the prostate in mr images by atlas matching.
\newblock In {\em 2007 4th IEEE International Symposium on Biomedical Imaging:
  From Nano to Macro}, pages 1300--1303, 2007.

\bibitem{reda_2016}
Islam Reda, Mohammed Elmogy, Ahmed Aboulfotouh, Marwa Ismail, Ayman El-Baz, and
  Robert Keynton.
\newblock Prostate segmentation using deformable model-based methods.
\newblock {\em Biomedical Image Segmentation: Advances and Trends}, page 293,
  2016.

\bibitem{Liu_2011}
Xin Liu, Masoom~A. Haider, and Imam~Samil Yetik.
\newblock Unsupervised 3d prostate segmentation based on diffusion-weighted
  imaging mri using active contour models with a shape prior.
\newblock {\em JECE}, 2011, jan 2011.

\bibitem{comelli_2021}
Albert Comelli, Navdeep Dahiya, Alessandro Stefano, Federica Vernuccio, Marzia
  Portoghese, Giuseppe Cutaia, Alberto Bruno, Giuseppe Salvaggio, and Anthony
  Yezzi.
\newblock Deep learning-based methods for prostate segmentation in magnetic
  resonance imaging.
\newblock {\em Applied Sciences}, 11(2):782, 2021.

\bibitem{Shen_2017}
D.~Shen, G.~Wu, and H-I. Suk.
\newblock Deep learning in medical image analysis.
\newblock {\em Annual Review of Biomedical Engineering}, 19(1):221--248, 2017.
\newblock PMID: 28301734.

\bibitem{Zhu_deeplycnn_2017}
Qikui Zhu, Bo~Du, Baris Turkbey, Peter~L. Choyke, and Pingkun Yan.
\newblock Deeply-supervised {CNN} for prostate segmentation.
\newblock {\em CoRR}, abs/1703.07523, 2017.

\bibitem{zabihollahy2019}
F.~Zabihollahy, N.~Schieda, S.~Krishna~Jeyaraj, and E.~Ukwatta.
\newblock Automated segmentation of prostate zonal anatomy on t2-weighted (t2w)
  and apparent diffusion coefficient (adc) map mr images using u-nets.
\newblock {\em Medical physics}, 46(7):3078--3090, 2019.

\bibitem{clark_2017}
T.~Clark, A.~Wong, M.~Haider, and F.~Khalvati.
\newblock Fully deep convolutional neural networks for segmentation of the
  prostate gland in diffusion-weighted mr images.
\newblock pages 97--104, 06 2017.

\bibitem{Rundo_usenet_2019}
Leonardo Rundo, Changhee Han, Yudai Nagano, Jin Zhang, Ryuichiro Hataya,
  Carmelo Militello, Andrea Tangherloni, Marco~S. Nobile, Claudio Ferretti,
  Daniela Besozzi, Maria~Carla Gilardi, Salvatore Vitabile, Giancarlo Mauri,
  Hideki Nakayama, and Paolo Cazzaniga.
\newblock Use-net: incorporating squeeze-and-excitation blocks into u-net for
  prostate zonal segmentation of multi-institutional {MRI} datasets.
\newblock {\em CoRR}, abs/1904.08254, 2019.

\bibitem{Rundo_datasets_2019}
Leonardo Rundo, Changhee Han, Jin Zhang, Ryuichiro Hataya, Yudai Nagano,
  Carmelo Militello, Claudio Ferretti, Marco~S. Nobile, Andrea Tangherloni,
  Maria~Carla Gilardi, Salvatore Vitabile, Hideki Nakayama, and Giancarlo
  Mauri.
\newblock Cnn-based prostate zonal segmentation on t2-weighted {MR} images: {A}
  cross-dataset study.
\newblock {\em CoRR}, abs/1903.12571, 2019.

\bibitem{Att-denseunet-2019}
Shuyi Li, Min Dong, Guangming Du, and Xiaomin Mu.
\newblock Attention dense-u-net for automatic breast mass segmentation in
  digital mammogram.
\newblock {\em IEEE Access}, 7:59037--59047, 2019.

\bibitem{SAUNetSA}
Changlu Guo, Marton Szemenyei, Yugen Yi, Wenle Wang, Buer Chen, and Changqi
  Fan.
\newblock Sa-unet: Spatial attention u-net for retinal vessel segmentation.
\newblock {\em ArXiv}, abs/2004.03696, 2020.

\bibitem{augmentor}
Marcus~D Bloice, Peter~M Roth, and Andreas Holzinger.
\newblock {Biomedical image augmentation using Augmentor}.
\newblock {\em Bioinformatics}, 35(21):4522--4524, 04 2019.

\bibitem{r2unet}
Md.~Zahangir Alom, Mahmudul Hasan, Chris Yakopcic, Tarek~M. Taha, and
  Vijayan~K. Asari.
\newblock Recurrent residual convolutional neural network based on u-net
  (r2u-net) for medical image segmentation.
\newblock {\em CoRR}, abs/1802.06955, 2018.

\bibitem{AttUnet}
Ozan Oktay, Jo~Schlemper, Loic~Le Folgoc, Matthew Lee, Mattias Heinrich,
  Kazunari Misawa, Kensaku Mori, Steven McDonagh, Nils~Y Hammerla, Bernhard
  Kainz, et~al.
\newblock Attention u-net: Learning where to look for the pancreas.
\newblock {\em arXiv preprint arXiv:1804.03999}, 2018.

\bibitem{taha2015}
A-A. Taha and A.~Hanbury.
\newblock Metrics for evaluating 3d medical image segmentation: analysis,
  selection, and tool.
\newblock {\em BMC medical imaging}, 15(1):1--28, 2015.

\bibitem{Yeung_2021}
Michael Yeung, Evis Sala, Carola-Bibiane Schönlieb, and Leonardo Rundo.
\newblock Unified focal loss: Generalising dice and cross entropy-based losses
  to handle class imbalanced medical image segmentation, 2021.

\bibitem{keras-unet-collection}
Yingkai Sha.
\newblock Keras-unet-collection.
\newblock \url{https://github.com/yingkaisha/keras-unet-collection}, 2021.

\bibitem{pytorch_implementation}
Lee~Jun Hyun.
\newblock Pytorch implementation of u-net, r2u-net, attention u-net, attention
  r2u-net.
\newblock \url{https://github.com/LeeJunHyun/Image_Segmentation}, 2019.

\end{thebibliography}

\end{document}